# Interlayer Coupling and Exciton Dynamics in 2D Hybrid Structures based on an InGaN Quantum Well coupled to a MoSe$_2$ Monolayer

D. Chen[1], D. Lagarde[2], L. Hemmen[2], L. Lombez[2], P. Renucci[2], M. Mauguet[2], L. Ren[2], C. Robert[2], N. Grandjean[1] and X. Marie[2,3*]

[1]Laboratory of Advanced Semiconductors for Photonics and Electronics, Institute of Physics, Ecole Polytechnique Fédérale de Lausanne (EPFL), CH-1015 Lausanne, Switzerland
[2]Université de Toulouse, INSA-CNRS-UPS, LPCNO, 135 Avenue Rangueil, 31077 Toulouse, France
[3]Institut Universitaire de France, 75231 Paris, France
* marie@insa-toulouse.fr

*Hybrid structures incorporating both III-nitride and Transition Metal Dichalcogenide (TMD) semiconductors have strong application potential for light harvesting and optoelectronics. Here we have investigated the properties of hybrid structures based on a MoSe$_2$ monolayer coupled to an InGaN quantum well (QW). The coupling efficiency is controlled by a thin GaN barrier of variable thickness located between them. Time-integrated and time-resolved micro-photoluminescence experiments show a quenching of the InGaN QW exciton emission which increases with the decrease of the GaN barrier thickness d: the PL intensity is reduced by a factor 3 for d=1 nm as a consequence of carrier transfer to the MoSe$_2$ monolayer. This interplay between the two semiconductors is confirmed by time-resolved photoluminescence spectroscopy highlighting a clear reduction of the QW exciton lifetime in the presence of the monolayer. Interestingly the coupling between the QW and the TMD monolayer is also demonstrated by measuring optically the excitonic transport properties in the quantum well: the exciton diffusion length decreases in the presence of the MoSe$_2$ monolayer. The measured dependences as a function of temperature highlight the role played by localization effects in the QW. All these results can be well interpreted by a type II band alignment between the InGaN QW and the MoSe$_2$ monolayer and a tunneling process between the two semiconductors.*

## I. Introduction

The development of III-nitride semiconductors has had a spectacular impact on the production of high-performance electronic and optoelectronic devices [1,2]. Atomically thin layers of semiconductor transition-metal dichalcogenides (TMDs) open up new possibilities for investigating two-dimensional (2D) physics and for potential new applications [3–5]. TMDs such as $MoS_2$, $WS_2$, $WSe_2$, $MoSe_2$, and $MoTe_2$ are direct band-gap semiconductors when thinned down to one monolayer and are characterized by very large exciton binding energies and unique spin/valley properties [6,7]. Combining these 2D TMD layered materials characterized by very large optical absorption coefficients with thin III-nitride compounds could broaden the scope of application and the performance of new hybrid devices.

Recent studies evidenced a type II band alignment between GaN and the TMD layer yielding an ultra-fast electron-hole separation following exciton photogeneration, with electrons accumulating on the GaN side and holes transferring to the TMD layer [8–10]. The surface passivation by $MoS_2$ coating was demonstrated recently in GaN/AlGaN quantum wells [11]. The electrostatic modulation of indirect excitons density thanks to few-layer graphene was also observed in GaN/(Al,Ga)N QWs [12].

The integration of atomically-thin TMD material with GaN presents strong potential for applications such as photo-catalysis, wideband photodetectors or even gas sensors [13–15]. However very little is known about the interplay between the two semiconductor layers.

In this work we demonstrate that the exciton properties of an InGaN QW can be engineered thanks to a proximity effect with a $MoSe_2$ monolayer. Remarkably the QW exciton lifetimes and diffusion lengths can be controlled as a result of hole transfer, from InGaN to $MoSe_2$ through a thin GaN barrier. The dependence of the dynamics as a function of the GaN barrier thickness allows us to rules out an interpretation based on energy transfer and the results can be understood on the basis of carrier tunneling from the InGaN QW to the $MoSe_2$ monolayer.

## II. Samples and experimental set-ups

The III-nitride samples are grown by metalorganic vapor phase epitaxy on c-axis free-standing GaN substrate (Fig. 1a). Starting from the substrate, the different layers include a 1 μm GaN epilayer, a superlattice (SL) composed of 24 pairs $In_{0.17}Al_{0.83}N$/GaN, a 21 nm GaN spacer, a 2.4 nm $In_{0.15}Ga_{0.85}N$ QW and finally a top-GaN barrier characterized by a thickness d. Three samples have been grown with d=1, 5 and 48 nm. We also study a reference sample with the same layer sequence except that there is no quantum well (no QW). The detailed growth conditions of these samples are provided in Ref. [16] for a similar structure. The hybrid TMD/GaN structures are obtained by deposition of a $MoSe_2$ monolayer (ML) on to the III-nitride samples described above using the widely used dry-stamping exfoliation technique in a nitrogen-regulated glovebox [17,18]. In order to protect the surface and avoid photo-doping effects [19], a thin hexagonal boron nitride (hBN) layer (~10-15 nm) is deposited on top of the $MoSe_2$ ML (the $MoSe_2$ bulk material is purchased at 2D Semiconductors and the hBN crystals are provided by NIMS-Tsukuba, Japan). Thermal annealing at 150°C for 10 min was performed after each transfer step to provide high-quality interfaces without bubbles nor residues [20]. The final structure is sketched in Fig. 1a and Figure 1b presents an optical microscopy image of the d=5 nm structure (top view), showing the $MoSe_2$ ML flake and the top hBN layer. It is worth noting that, at low temperature, the QW emission from the d=1 and 5 nm samples remains sharp and well-defined even after $MoSe_2$ deposition, with no noticeable spectral broadening. This observation suggests that the $MoSe_2$/GaN interface is of high quality, as any significant interfacial degradation would likely affect the emission properties of these surface-sensitive QWs.

Micro-Photoluminescence (PL) experiments are used to record the emission spectra of the different samples in the temperature range T=10–300 K. The samples are placed on three-axis stepper motors to control the sample position with nm precision inside a low-vibration closed cycle He cryostat [21]. This allows us to easily investigate the optical properties of the InGaN/GaN QW with (w/) or without (w/o) the $MoSe_2$ monolayer on the top. For the investigation of the optical properties of the $MoSe_2$ ML alone, a He-Ne laser ($\lambda$=633 nm) is used. For the other PL measurements, the samples are excited by a frequency-doubled mode-locked Ti:sapphire oscillator with ~1.5 ps pulses with a repetition rate of 800 kHz ($\lambda$=405 nm). The time-integrated PL spectra are recorded by a cooled charged-coupled device camera coupled to a spectrometer. The time-resolved spectra and kinetics are recorded with a Hamamatsu streak camera in the triggered mode with an overall instrument response time resolution (IRF) of ~1 ns [22]. The typical size of the excitation/detection spot in the micro-PL experiments is 1 $\mu$m. The white light source for the differential reflectivity measurements is a halogen lamp with a stabilized power supply. For the exciton diffusion measurements, a continuous wave (cw) laser with high quality spatial profile is used ($E_{laser}$=3.306 eV) together with a high sensitivity CMOS 2D camera.

**III Results and discussions**

*InGaN QW optical properties*
Figure 1c displays the PL spectrum of the InGaN QW on the d=5 nm sample at low temperature (T=10 K). It is dominated by a strong optical recombination of the ground state exciton (E=2.92 eV) and a weaker phonon-assisted emission (E=2.83 eV). Despite the high quality of the structure (attested by a very bright PL intensity), the exciton PL linewidth is about 27 meV (Full Width at Half Maximum, FWHM), indicating a significant inhomogeneous broadening. The variation of the exciton PL peak energy as a function of temperature is presented in Fig. 1d. The well-known S-shape behavior of the peak energy (with a local minimum at ~60 K) is clearly observed for temperatures smaller than 150 K, as a consequence of exciton localization on inhomogeneous potential fluctuations [23].

*$MoSe_2$ monolayer optical properties.*
Figure 2 displays the low temperature cw photoluminescence spectra of the $MoSe_2$ monolayer for d=1, 5 and 48 nm, as well as in the absence of the QW ("no QW"). We emphasize that the He-Ne laser energy is E=1.96 eV such that carriers are only photogenerated in the $MoSe_2$ ML since the laser energy is much smaller than the InGaN QW gap energy. For all samples the luminescence is dominated by two peaks corresponding to the recombination of neutral exciton ($X_0$) and charged exciton (trion, T) as observed by many groups [24–26]. The black line in Fig. 2b displays the differential reflectivity spectrum, confirming the assignment of the two peaks to the neutral exciton and trion lines (the neutral exciton peak is always stronger in absorption/reflectivity spectra for residual doping as it is the case here) [27]. We have recorded different spectra obtained for different positions on the $MoSe_2$ ML flake. We observe slight variations in terms of PL peak intensities or broadening from point to point but overall, the results are quite homogeneous depending on the point of the TMD sample studied. We note that for the "no QW" and the d=48 nm samples, the luminescence linewidths of the $X_0$ and T lines are typically 3 and 5 meV (FWHM) respectively. These narrow linewidths are state-of-the-art and reveals the high quality of the monolayers deposited on the GaN barrier. These widths are comparable to those obtained with encapsulation using hexagonal boron nitride which is the material of choice to control the dielectric environment of 2D materials [18,28,29]; however the trion to neutral exciton PL intensity ratio in Fig. 2 is larger than the one observed in hBN fully encapsulated TMD monolayers , likely due to charge transfer from the GaN surface.

*Interplay between the InGaN QW and the MoSe$_2$ monolayer*
Figure 3a displays the time-integrated photoluminescence intensity of the InGaN QW in the d=5 nm structure using a laser excitation energy above the QW gap but below that of the other III-nitride layers in the structure ($E_{laser}$=3.06 eV). Here the laser excitation power is low ($P_{laser}$=0.5 μW). The striking feature is a significant decrease of the luminescence intensity in the presence of the MoSe$_2$ monolayer on top (compare the curve with or without MoSe$_2$ ML, obtained by translating the sample, Fig. 1a). It is even more pronounced in the d=1 nm structure since the intensity decreases by a factor ~3 when the MoSe$_2$ ML is present ($P_{laser}$=0.5 μW). We define the PL quenching factor as: $Q=I_{w/o}/I_{w/}$, where $I_{w/o}$ ($I_{w/}$) denotes the total QW PL intensity without (with) the MoSe$_2$ monolayer. For the barrier thickness d=5 nm, we find Q ~2, whereas for the d=48 nm structure, the PL intensity is very similar (Q ~1.2) if the monolayer is present or not (see Fig. 3b). For d=48 nm, one notes a slight decrease when the 2D semiconductor is present on the top; it can be easily explained by both the absorption of laser and photons emitted by the QW when they travel across the MoSe$_2$ ML. However, we emphasize that the variation of Q displayed in Fig. 3b cannot be explained by the optical absorption through the ML since it is identical for d=1, 5 and 48 nm. These results evidence a clear electronic interplay between the InGaN QW and the MoSe$_2$ ML. The luminescence quenching becomes more pronounced with thinner GaN barriers, suggesting a distance-dependent coupling between the QW and the 2D layer, possibly involving carrier transfer through mechanisms such as tunneling.

Figure 4 presents the key results of this work. In order to confirm the carrier transfer from the QW to the 2D monolayer, we have measured the InGaN QW exciton dynamics by time-resolved photoluminescence for the different thicknesses of the GaN barrier d=1, 5 and 48 nm. For each thickness, the kinetics have been recorded with or without the presence of the MoSe$_2$ monolayer on top of the structure (simply by translating the sample, see Fig. 1a). As expected, the exciton dynamics is identical for the thick barrier d=48 nm (left panel in Fig. 4a). No carrier transfer can occur between the QW and the MoSe$_2$ ML for such a large thickness. In contrast, for d=5 and 1 nm, the QW luminescence decay time is clearly shorter in the presence of the MoSe$_2$ monolayer and it decreases with d (middle and right panels in Fig. 4a). Figure 5 highlights the spectral dependence of the coupling between the QW and the 2D ML: the QW dynamics is altered by the presence of the MoSe$_2$ ML mainly on the high energy side of the QW spectrum. Remarkably the modification of the InGaN QW exciton dynamics induced by the presence of the MoSe$_2$ monolayer is observed up to room temperature, as shown in Fig. 6 for the d=5 nm hybrid structure (similar results have been obtained for the d=1 nm sample).

Interestingly the coupling between the QW and the TMD monolayer can also be demonstrated by measuring the excitonic transport properties in the QW. This exciton transport can be probed using a simple photoluminescence experiment based on a diffraction-limited laser excitation that induces lateral diffusion of the photogenerated excitons [30–33]. Figure 7a displays the PL image at 300 K without the MoSe$_2$ ML for the d=5 nm hybrid structure (the blue circle corresponds to the image of the diffraction-limited laser spot). The excitonic luminescence comes from a spatial region which extends significantly beyond the excitation spot, as a consequence of exciton diffusion [31]. Remarkably we observe that the spatial extension of the luminescence is smaller in the presence of the MoSe$_2$ monolayer and the difference between the w/ and w/o curves is larger if d decreases (Fig. 7b, T=300 K). This is a direct consequence of the coupling between the InGaN QW and the MoSe$_2$ ML, yielding a reduction of the QW exciton diffusion length due to a decrease of the exciton lifetime, as shown in Fig. 4 and Fig. 6. Let us recall that the diffusion length writes simply $L = \sqrt{D\tau}$, where τ and D are the exciton lifetime and diffusion coefficient, respectively. We do not expect a variation of the latter with a monolayer located 5 nm from the QW. The PL profiles can be fitted taking into account the convolution with the measured laser's gaussian profile and using the procedure described in Ref. [30], where the diffusion length L is the only adjustable parameter. For the QW with d=5 nm (without MoSe$_2$), $L$ is approximately 700 nm at room temperature (Fig. 7c), and the exciton lifetime at 275 K is about 18 ns (Fig. 6d), resulting in $D = 0.27 \text{ cm}^2\text{s}^{-1}$. This diffusion coefficient is in good agreement with previously reported value [34]. Figure 7c presents the variation of the measured diffusion length with or without MoSe$_2$ ML as a function of temperature. The reduction of L in the presence of the MoSe$_2$ ML is clearly observed for temperatures larger than > ~100 K. For lower temperatures, the diffusion coefficient and hence the diffusion length drop due to localization effects (Fig. 1d). This localization prevents the diffusion of excitons on long distance. In these conditions our experimental set-up with limited spatial resolution does not allow to highlight a difference of the exciton diffusion characteristics with or without MoSe$_2$.

*Discussion on the coupling between the InGaN quantum well and the MoSe$_2$ monolayer*
A detailed analysis of the experimental results raises several remarks.
First, on time-resolved PL traces presented in Fig. 4a, we note rather long lifetimes with two regimes likely due to recombination dynamics of free and localized excitons rather than Auger-type recombination effects as they do not depend on the excitation power [35]. Moreover, we believe that the internal electric field in the 2.4 nm-thick InGaN QW plays a minor role here. This is confirmed by measuring the energy shift of the time-resolved QW PL spectra (not shown), which is the same under high or low excitation power [36]. We have obtained very similar results with an InGaN QW thickness of 0.8 nm (not shown), where the Stark effects are negligible. In Fig. 5b, for the d=5nm structure, we observe that the difference in the InGaN QW exciton kinetics in the presence or absence of the MoSe$_2$ monolayer is much more marked for higher energy excitons: the decay time is shorter for exciton energies in the high energy part of the spectrum compared to the one measured at lower energies. This demonstrates that the efficiency of the coupling between the InGaN QW and the MoSe$_2$ ML decreases as a consequence of the carriers' localization due to potential fluctuations [37].
Second, we note that the decay times for long delays are very similar with or without MoSe$_2$ ML (see Fig. 4a for t>~30 ns and Fig. 5b), though the initial decay times are very different; this is the case for both d=1 nm and d=5 nm samples. This phenomenon is interpreted again as a consequence of the localization of carriers over time in the InGaN QW, which progressively suppresses the carrier transfer. This is consistent with both the PL inhomogeneous broadening and the PL S-shape behavior presented in Fig. 1d. We believe that exciton localization effects also explain the variation of the PL quenching factor with excitation power (Fig. 3b). Indeed, we observe that the Q factor increases with excitation power for d=1 and d=5 nm structures (whereas it is constant within the experimental accuracy for d=48 nm). For the d=5 nm structure, the PL quenching factor increases from Q=2 for $P_{laser}$=0.5 µW to Q=3 for $P_{laser}$= 5 µW. For low photogenerated exciton densities, the localization effect will limit the transfer to the MoSe$_2$ ML. For larger densities, the lower energy exciton states will be saturated and more efficient transfer will occur through higher energy exciton states with larger spatial extension. In order to extract quantitative information, we have considered the following simple model. We use rate equations to characterize the evolution of the hot ($N_{hot}$), free ($N_{free}$) and localized ($N_{loc}$) exciton states as a function of time. Our model, schematically illustrated in Fig. 4b, considers the generation of a hot QW exciton population after laser excitation, its relaxation to free (cold) exciton states with a characteristic decay time $\tau_{e_1}$. The free exciton population can either relax to localized states with a characteristic time $\tau_{e_2}$ or recombine with a decay time $\tau_1$. The localized exciton population is characterized by a decay time $\tau_2$. The possible charge transfer from the QW to the MoSe$_2$ ML is represented by a transfer time $\tau_{transfer}$ that can only affect the hot and free exciton populations. The rate equations are numerically solved using parameters presented in Fig. 4d and the results are presented in Fig. 4c for the different samples given that the total PL dynamics correspond to the recombination of both free and localized excitons. The fitting protocol is as follows: first, we fit the PL dynamics without the MoSe$_2$ ML ($\tau_{transfer}$ is infinite). This yields $\tau_{e1}$=0.75 ns limited by the IRF, $\tau_{e2}$=3.5 ns, $\tau_1$~10 ns and $\tau_2$~15 ns for the three samples. Then we fit the exciton dynamics of the InGaN QW in the presence of the MoSe$_2$ ML by considering a finite decay time $\tau_{transfer}$ corresponding to the transfer time from the QW to the MoSe$_2$ ML. Using the values of $\tau_1, \tau_2, \tau_{e1}$ and $\tau_{e2}$ previously determined, the exciton dynamics in the hybrid structure ("w/ MoSe$_2$") is well reproduced with $\tau_{transfer}$=6 ns in the d=5 nm sample (see middle panel in Fig. 4c). The same fitting protocol yields $\tau_{transfer}$=4 ns for the d=1 nm hybrid structure (see right panel in Fig. 4c and table in Fig. 4d). Remarkably, taking into account the free or the localized exciton population independently, we could reproduce with the same parameters the spectral dependence of the charge transfer from the InGaN QW to the 2D ML, as shown in Fig. 5. Despite the simplicity of the model, it catches the main features of the kinetics. Furthermore, since the quenching factor reflects the competition between carrier transfer and recombination rates, the fitted transfer rate can be directly correlated with the measured quenching parameters obtained under the same excitation conditions. It can be easy shown that $\frac{1}{\tau_{transfer}} \propto Q - 1$. For 5 µW excitation power, we measured $Q \approx 4.25$ for $d = 1$ nm and $Q \approx 3$ for $d = 5$ nm (Fig. 3b), leading to $\frac{\tau_{transfer}(d=5 \text{ nm})}{\tau_{transfer}(d=1 \text{ nm})}$~1.6.

This ratio is in very good agreement with the transfer times in Fig. 4d : $\frac{\tau_{transfer}(d=5 \text{ nm})}{\tau_{transfer}(d=1 \text{ nm})}$~1.5. Furthermore, using the value from Fig. 4d in Eq. (1), we deduce $\tau_0 \approx 13$ ns, consistent with $\tau_1$ and $\tau_2$ in the same fitting.

This consistency between the quenching factor and the extracted transfer rate supports the reliability of our fitting approach and confirms the coherence of the overall data analysis.

The drop of the transfer time from $\tau_{transfer}$ =6 ns for d=5 nm to $\tau_{transfer}$ =4 ns for d=1 nm clearly evidences that the coupling between the QW and the MoSe$_2$ ML increases when the barrier thickness decreases. It also rules out an interpretation based on an energy transfer process between the QW and the TMD monolayer (induced by electron exchange or dipole-dipole interaction) [38–40]. It is well known that energy transfer depends critically on the donor-acceptor distance, the band offsets and the dielectric environment. For instance, Förster-like energy transfer has a typical scaling law $d^{-n}$, where n is determined by the dimensionality (n=2 for instance for a 2D/2D structure) [41,42]. The dependence of $\tau_{transfer}$ on the barrier thickness cannot therefore be explained by an energy transfer process.

Overall, we interpret the reduction of the InGaN QW exciton lifetime in the presence of the monolayer by a carrier transfer from the quantum well to the monolayer. The presence of the MoSe$_2$ ML can lead to exciton dissociation due to the type II alignment, leading to charge transfer, thereby influencing the exciton lifetime and excitonic transport properties [13]. This can occur thanks to a defect-assisted tunneling process which was evidenced previously in InGaN/GaN QW structures [43,44]. Based on an experimental and numerical investigations, an efficient defect-assisted tunnel effect through a few nm GaN barrier layer was for instance interpreted as an important contribution to the current flow below the optical turn-on of LEDs [44]. However, this remains in our case a plausible yet unconfirmed mechanism. The transferred charges may either be captured by surface states at the MoSe$_2$/GaN interface or contribute to trion formation in the ML. Both processes involve exciton dissociation in the QW, followed by charge transfer to the interface induced by the presence of the ML, and contribute to the additional non-radiative recombination observed in the surface QWs (d=1 and 5 nm). Distinguishing between these processes requires further study of the electronic band structure of the two materials and their surface states near the interface.

Finally, we emphasize that we have probed the coupling between the QW and the TMD monolayer by measuring the change of the QW exciton properties (lifetime, diffusion length). One may ask whether it is possible to also see a signature of the coupling on the emission properties of the MoSe$_2$ monolayer . In the excitation conditions of Fig. 3 and Fig. 4, the carriers are photo-generated both in the InGaN QW and in the MoSe$_2$ ML. So, we can also record the exciton luminescence dynamics of the MoSe$_2$ ML for the same laser excitation energy. Figure 8 presents both the InGaN QW exciton and the MoSe$_2$ ML exciton dynamics for the d=5 nm hybrid structure. The MoSe$_2$ ML dynamics is characterized by a very short decay time (less than 1 ns) controlled by our instrument time-resolution; it is followed by a second regime characterized by a weak intensity and much longer decay time (tens of ns) which is very similar to the InGaN QW one. The first decay results directly from the very short (~1 ps) exciton lifetime of MoSe$_2$ ML as a consequence of its very large binding energy [45,46]. Note that bare MoSe$_2$ MLs do not exhibit a second long decay time [47] . The MoSe$_2$ longer luminescence decay in Fig. 8 (on the tens-of-nanoseconds scale), observed whatever the barrier thickness, is unlikely to originate from energy or carrier transfer from the QW to the ML, considering (i) the measured weak d-dependence of this decay (not shown) and (ii) the type-II band alignment between the two materials. Instead, it results likely from re-absorption of photons emitted by the InGaN QW, a process distinct from direct energy or charge transfer, which would naturally produce long decay dynamics in the ML, similar to those observed in the QW.

In summary, we have measured the exciton dynamics and diffusion properties in a hybrid structure made of an InGaN QW coupled to a MoSe$_2$ monolayer via a thin tunnel barrier. The results can be well interpreted by carrier tunneling from the InGaN quantum well to the MoSe$_2$ monolayer. The control of QW exciton lifetime and diffusion length induced by proximity effect with a 2D material could offer new perspectives for optoelectronic applications.

**Acknowledgments**: We are grateful to T. Taniguchi and K. Watanabe (NIMS Tsukuba) for providing us high quality hBN crystals. We thank Jean-François Carlin for the MBE growth and K. Mourzidis for the fabrication of one sample. This work was supported by the Agence Nationale de la Recherche under the program ESR/EquipEx+ (Grant No. ANR-21-ESRE- 0025) and the Swiss National Science Foundation through Grant 200021E 175652.

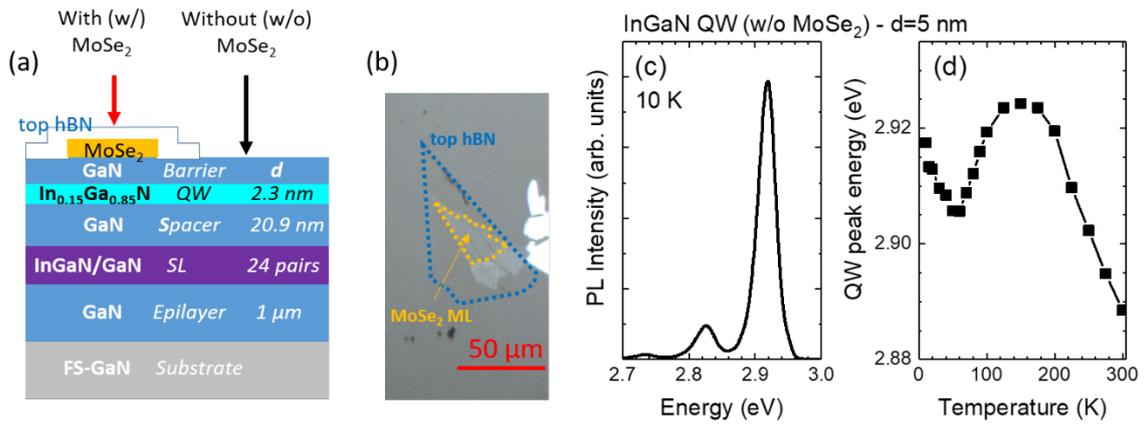

Fig 1: (a) Schematics of the sample structure highlighting the two excitation conditions: with (red arrow) or without (black arrow) MoSe$_2$ ML on top of the InGaN QW. (b) Optical microscopy image of hBN-encapsulated MoSe$_2$ ML deposited on the surface of the InGaN/GaN QW with d=5 nm (top view). (c) PL spectrum of the InGaN QW of the d=5 nm sample at low temperature (T=10 K), P$_{exc}$=0.5 µW@800 kHz. (d) Dependence of the exciton PL peak energy as a function of temperature.

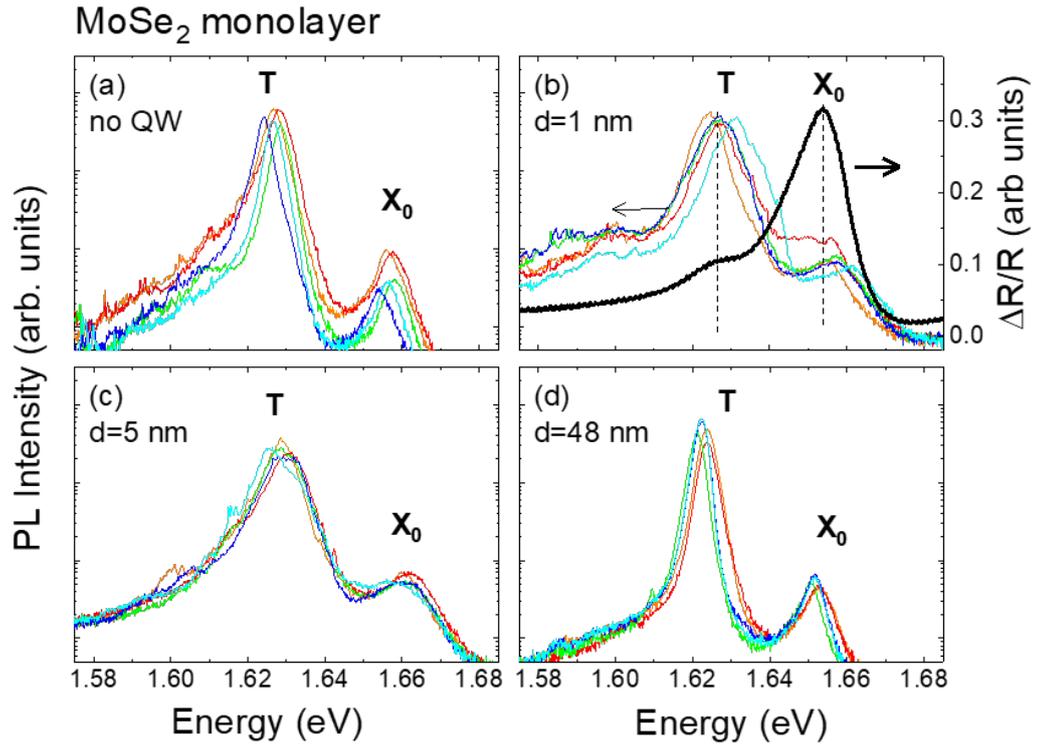

Fig. 2: (a)-(d) cw photoluminescence spectra of the MoSe$_2$ monolayer for d=1, 5 and 48 nm, as well as in the absence of the QW ("no QW") measured at T=10 K, P$_{laser}$=5 µW, E$_{laser}$=1.96 eV. The different colored spectra correspond to different positions on the MoSe$_2$ ML flakes. Black curve in (b) also displays the differential reflectivity spectrum.

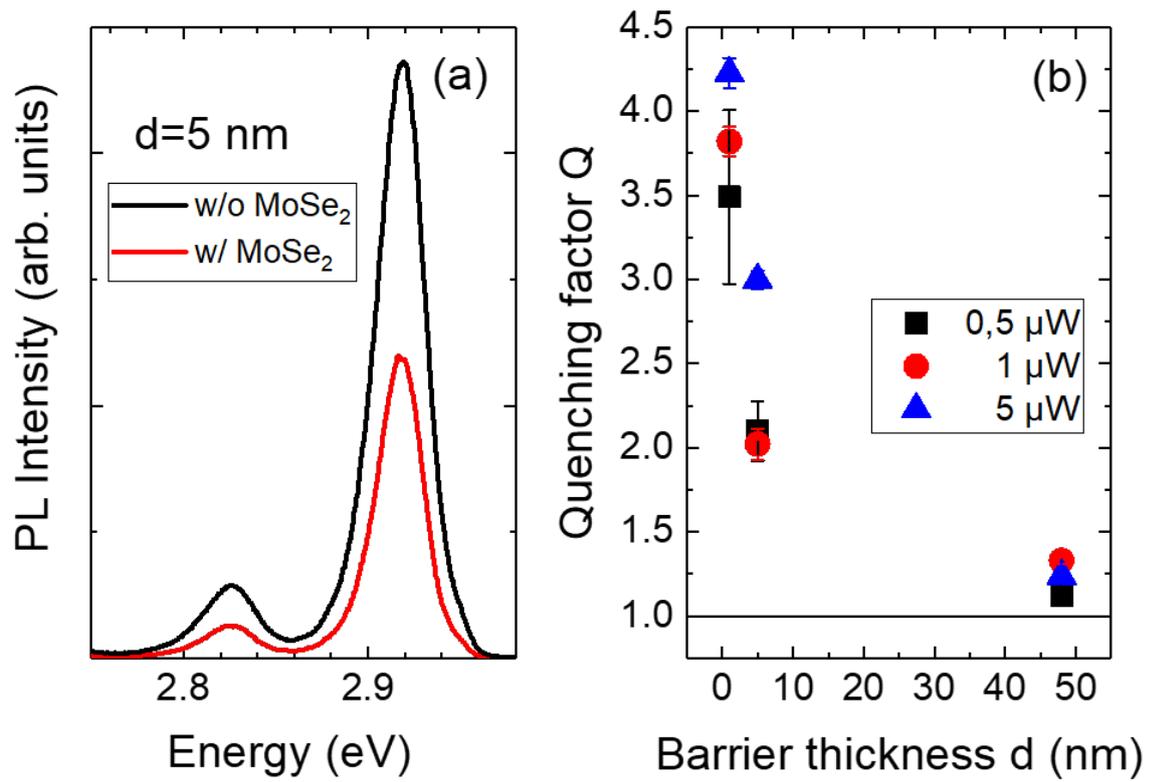

Fig. 3: (a) Time-integrated PL spectrum of the InGaN QW in the d=5 nm structure at T=10 K, with (w/) and without (w/o) MoSe$_2$, (b) Evolution of PL quenching factor Q as a function of the barrier thickness d for different excitation powers.

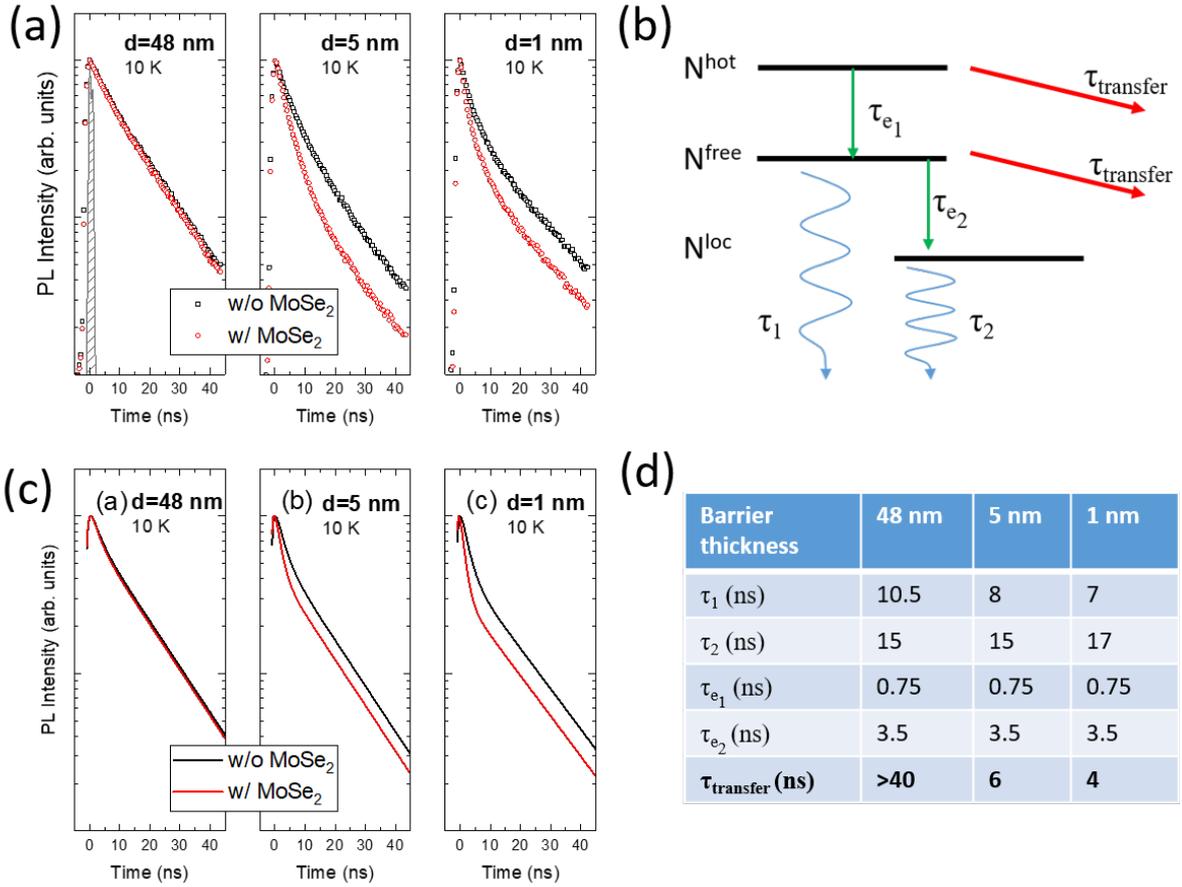

Fig. 4: (a) Normalized PL dynamics of the InGaN QW exciton with different GaN surface barrier thicknesses, measured on regions without (black curves) or with (red curves) the MoSe$_2$ ML, T=10 K, P$_{exc}$=5 µW@800 kHz; the kinetics are taken on the spectrally integrated exciton line. The hatched grey curve on left panel is the IRF. (b) Schematic representation of the exciton energy levels and relaxation channels considered by the fitting model (see text) (c) Normalized fitted QW PL dynamics using the parameters presented in (d) for the different samples without (black curves) or with (red curves) top MoSe$_2$ ML.

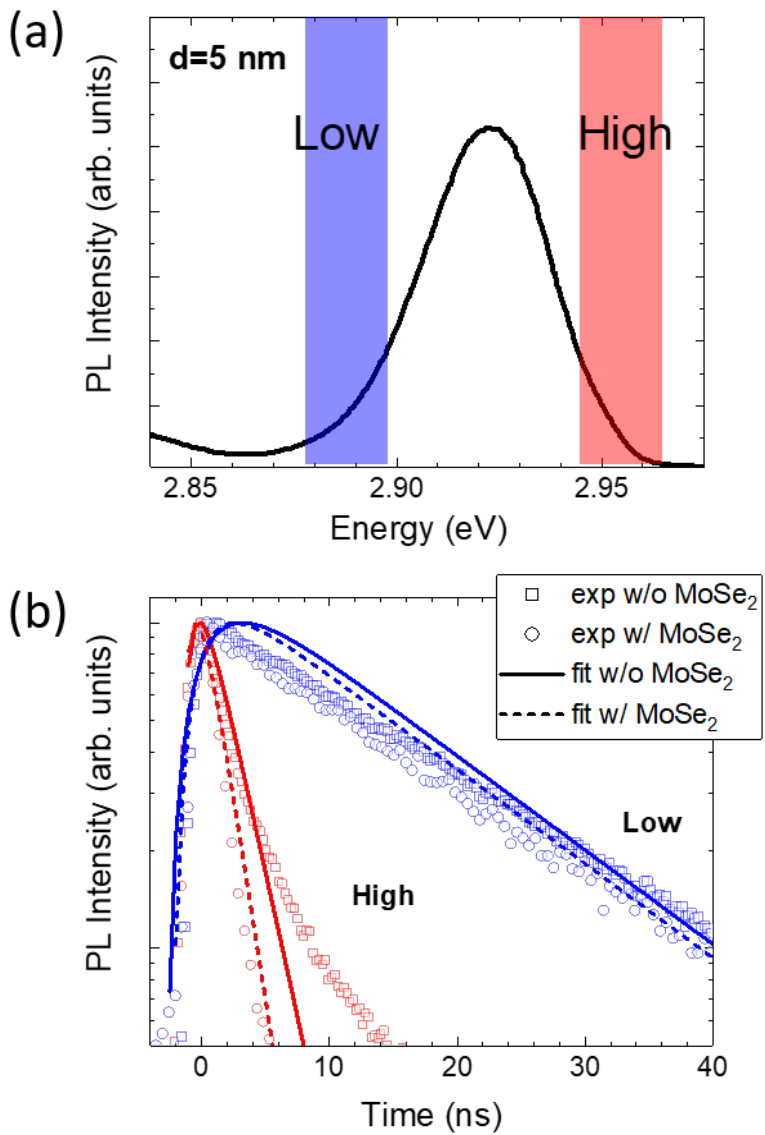

Fig. 5: (a) PL spectrum of the InGaN QW with d=5 nm measured at 10 K, $P_{exc}$=5 µW@800 kHz. (b) Normalized PL dynamics of the InGaN QW spectrally-integrated in energy ranges presented in (a): blue (red) traces are taken at the low (high) energy part of the QW spectrum. Fitted normalized PL dynamics are also presented considering only the "free" (red) and "localized" (blue) QW exciton intensity on sample position without (solid lines) and with (dashed lines) $MoSe_2$ ML.

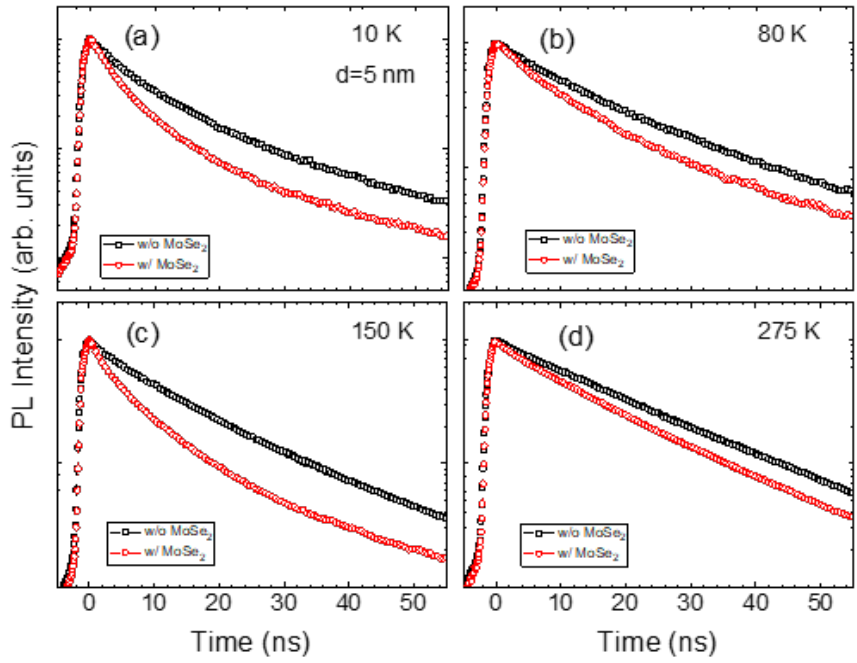

Fig. 6: Normalized PL dynamics of the InGaN QW for the d=5 nm sample, measured on regions w/o (black curves) or w/ (red curves) MoSe2 ML, for different temperatures: (a) 10 K, (b) 80 K, (c) 150 K and (d) 275 K.

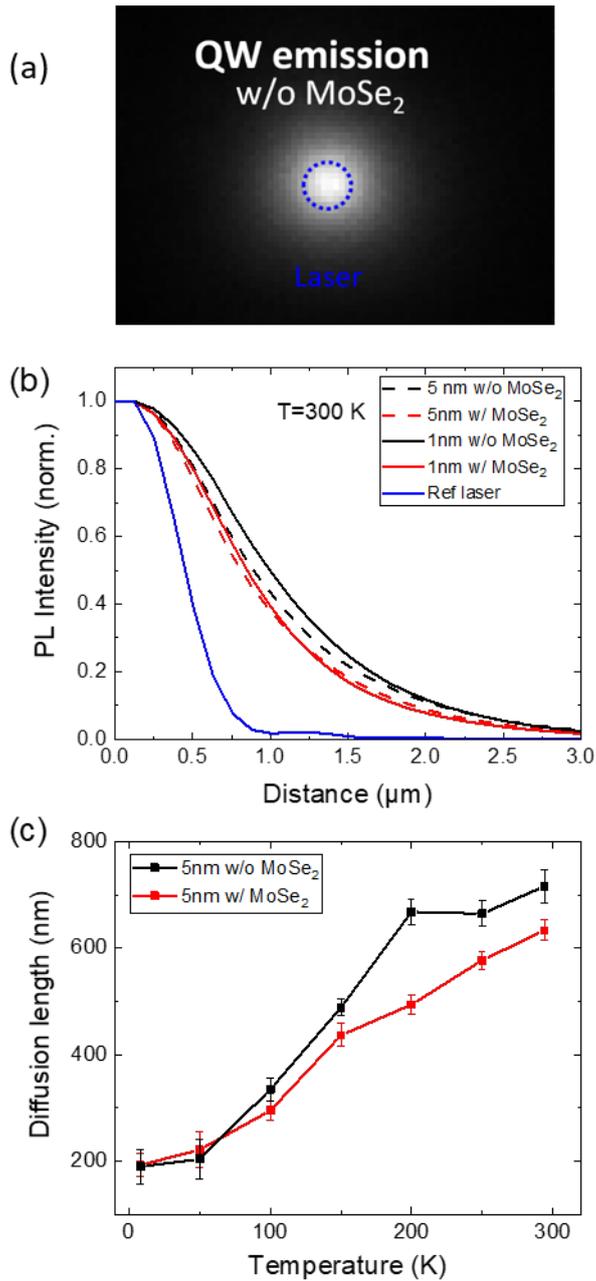

Fig. 7: (a) PL image of the QW emission at 300 K without the MoSe$_2$ ML for the d=5 nm sample (plotted with an intensity scale ranging from 0 (black) to 1 (white) ). The blue circle corresponds to the image of the diffraction-limited laser spot. (b) Spatial profile of the QW luminescence without (w/o, black curves) or with (w/, red curves) MoSe$_2$ ML. The blue curve corresponds to the excitation laser profile. (c) Evolution of the measured diffusion length without (black) or with (red) MoSe$_2$ ML as a function of temperature.

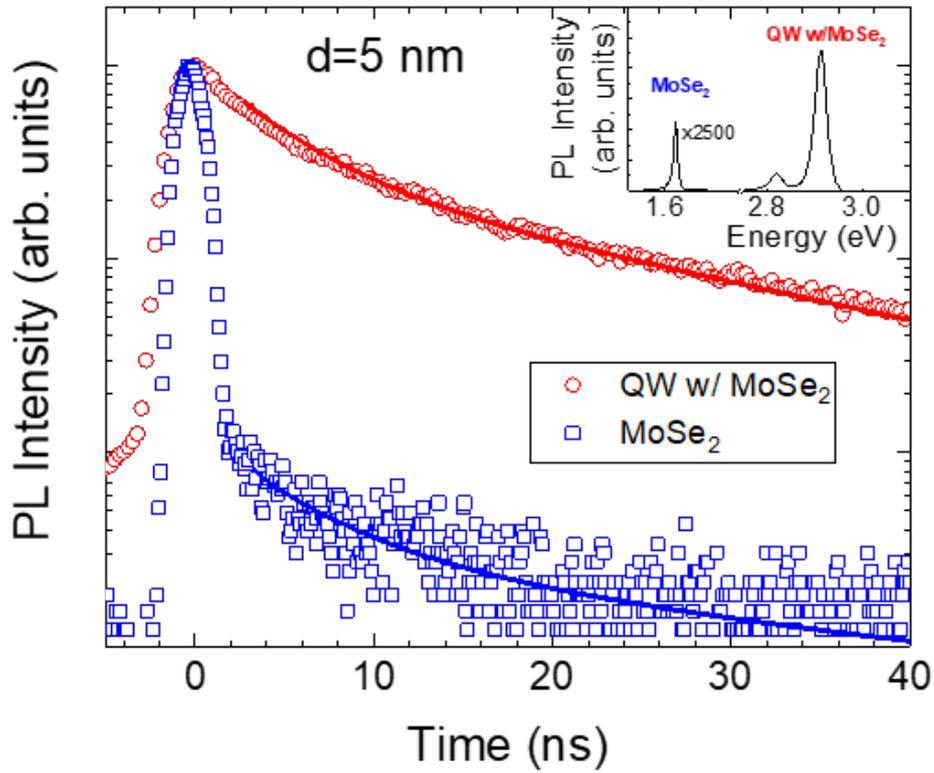

Fig. 8: Normalized PL dynamics of InGaN QW exciton (red) and MoSe$_2$ ML exciton (blue) for the d=5 nm hybrid structure. The lines are guides for the eye. Inset: Time-integrated PL spectra of the QW and MoSe$_2$ ; T=10 K, P$_{exc}$=5 µW@800 kHz and E$_{laser}$=3.06 eV.